\documentclass[11pt]{article}
\usepackage{amsmath}
\usepackage{amssymb}
\usepackage{amsfonts}
\usepackage{amscd}
\usepackage{accents}
\usepackage{pst-all}
\usepackage{epsfig}
\usepackage{graphicx}
\usepackage[tight]{subfigure}
\date{\today}
\newcommand{\bmat}{\left(\begin{array}}
\newcommand{\emat}{\end{array}\right)}
\newcommand{\be}{\begin{equation}}
\newcommand{\ee}{\end{equation}}
\newcommand{\ba}{\begin{eqnarray}}
\newcommand{\ea}{\end{eqnarray}}
\newcommand{\uone}{U(1)}
\newcommand{\uonep}{U(1)$^\prime$}
\newcommand{\sunp}{SU(N)$^\prime$}
\newcommand{\sunpm}{SU(N$-$1)$^\prime$}

\def\lsim{\raise0.3ex\hbox{$\;<$\kern-0.75em\raise-1.1ex\hbox{$\sim\;$}}}
\def\gsim{\raise0.3ex\hbox{$\;>$\kern-0.75em\raise-1.1ex\hbox{$\sim\;$}}}

\def\be{\beta}

\begin{document}

\vspace*{-.6in} \thispagestyle{empty}
\begin{flushright}
IFT-UAM/CSIC-15-078
\end{flushright}
\baselineskip = 20pt

\vspace{.5in} {\Large
\begin{center}
{\bf  Higgsophilic gauge bosons and monojets at the LHC 
  }
\end{center}}

\vspace{.5in}

\begin{center}
{\bf Jong Soo Kim$^1$, Oleg Lebedev$^2$ and  Daniel Schmeier$^3$    }  \\

\vspace{.5in}

$^1$\emph{Instituto de F\'{i}sica Te\'{o}rica  UAM/CSIC, Madrid, Spain
}\\
$^2$\emph{Department of Physics and Helsinki Institute of Physics,
Gustaf H\"allstr\"omin katu 2, FIN-00014 University of Helsinki, Finland
 }\\
$^3$\emph{Physikalisches
  Institut and Bethe
  Center for Theoretical Physics, University of Bonn, Bonn, Germany
 }  
\end{center}

\vspace{.5in}

\begin{abstract}
\noindent
 We consider a generic  framework where the Standard Model (SM) coexists with a hidden sector endowed with some additional gauge symmetry. When this symmetry is broken by a scalar field charged under the hidden gauge group, the corresponding scalar boson generally mixes with the SM Higgs boson. In addition, massive hidden gauge bosons emerge and via the mixing, the observed  Higgs--like mass eigenstate is the only known particle that couples to these hidden gauge bosons directly. 
 We study the LHC monojet signatures of this scenario and the corresponding constraints on the gauge coupling of the hidden gauge group as well as the mixing of the Higgs scalars.
 \end{abstract}

 \newpage
 
\section{Introduction}

When it comes to  singlet scalar extentions of the Standard Model (SM), the SM Higgs field $H$ plays a special role: Since $H^\dagger H$ is the only Lorentz and gauge invariant SM operator with mass dimension less than four, it is only the SM Higgs that can couple to extra SM singlet scalars $X$ at the renormalizable level \cite{Silveira:1985rk,Schabinger:2005ei,Patt:2006fw}. This has interesting implications for the Higgs evolution in the early Universe, for example if the extra singlet is an inflaton \cite{Gross:2015bea}.

\begin{figure}[h!]
  \centering
         \includegraphics[width=0.35\textwidth]{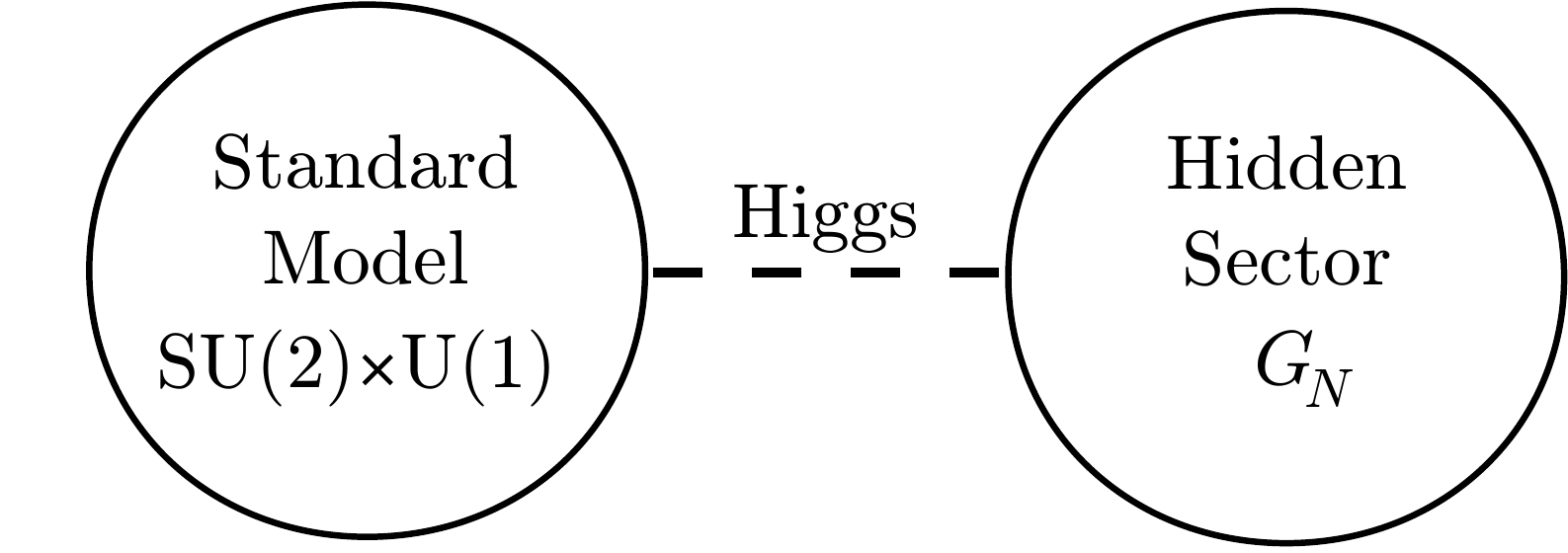}  
          \caption{Graphical illustration of the Higgs portal setup: A 'Hidden Sector' is assumed to be decoupled from the Standard Model, except for the Higgs boson which can couple both to the SU$(2)\times$U$(1)$ gauge sector of the Standard Model and the gauge group $G_N$ of the hidden sector.}
\label{1}
\end{figure}

Suppose that the extra singlet scalar $X$ is linked to a ``hidden sector'', which itself embeds a new gauge group $G_N$. If we are to probe the gauge bosons associated with this hidden gauge group, the feature  described above can be used to construct a ``Higgs Portal'': Here, $H$ is the only SM field that couples directly to these gauge bosons, as depicted schematically in Fig.\ref{1}. Na\"{\i}vely, one could give a hidden sector charge to $H$ to make it couple to the $G_N$ gauge bosons. However, in that case gauge invariance of the SM Yukawa couplings 
would require the SM fermions to be charged under $G_N$ as well. They would hence couple to the hidden gauge sector too and thus violate the assumption of a Higgs Portal. 

A viable alternative would instead be to let $X$ be charged under $G_N$ and mix with the SM $H$. In fact,  the Higgs portal coupling 
\begin{equation}
\Delta V = {\lambda_{hx}}~ H^\dagger H X^\dagger X
\end{equation}    
is renormalizable, Lorentz invariant and complies with both the hidden sector and the SM gauge symmetries. As such
it can and should be included in any theory describing the Standard Model and a hidden sector of the above type. 

Assuming that both $H$ and $X$ develop vacuum expectation values (vevs), this coupling automatically leads
to a mixing between the two fields upon spontaneous breaking of both symmetries. Therefore the observable mass eigenstates can couple to both the SM and the hidden sector. Most importantly, the recently observed SM-like Higgs boson can couple to the massive gauge bosons of the broken $G_N$, while all other SM particles cannot. We thus call the hidden sector gauge bosons ``Higgsophilic'' \footnote{The term ``Higgsophilic'' has first appeared in \cite{Fan:2011vw} in a different context.}. If they are themselves stable  or only decay into other  stable hidden sector fields, their production at colliders would  appear as missing net momentum in the event. In this work, we study a particular LHC signature of the hidden sector using this argument, namely the production of a single jet associated with sizable missing transverse momentum, also called ``monojet search''.

Previous studies of related models include  \cite{Gopalakrishna:2008dv}, where 
a kinetic mixing between the bosons associated to the SM \uone{} and a hidden sector \uonep{} boson was assumed. In that case however, the SM Higgs boson is not the only SM field that couples to the hidden sector which entails different collider signatures. LHC studies of a specific limit in which the second Higgs scalar decouples  \cite{Lebedev:2011iq}  was considered in \cite{Endo:2014cca},
in which case the obtained estimates are considerably optimistic (see also \cite{Djouadi:2011aa,Djouadi:2012zc}).  
A recent paper  \cite{Chen:2015dea} presents an interesting related LHC study focusing on the vector boson fusion (VBF) channel in a specific dark matter (DM) scenario and a different kinematic range than discussed here.

In our work, we consider a more general framework in which the heavy Higgs-scalar couples to arbitrary
gauge bosons of the hidden sector and we improve on previous studies in technical aspects. After introducing the model and its phenomenological implications in  Sec.~\ref{sec:zprime} we discuss the methodology and the results of our collider  study in Sec.~\ref{sec:monojet}.

\section{Models for Higgsophilic $Z^\prime$}
\label{sec:zprime}
\subsection{$Z^\prime$ from a  hidden \uonep}
\paragraph{Lagrangian, mass eigenstates and couplings:}
We outline the most important phenomenological results for our model below. For more details, see e.g. \cite{Lebedev:2011aq}.

Consider the Standard Model extended by a hidden sector with a  $sequestered$  \uonep{}, which by construction is 
orthogonal to the SM \uone. The hidden sector   contains the vector gauge field $A_\mu^\prime$ associated 
 with the \uonep{} and a complex scalar $X$ charged under the \uonep{} but neutral  under the SM gauge group.  The kinetic terms of these fields read
\begin{equation}
{\cal L}_{\rm kin} = -{1\over 4} F^\prime_{\mu\nu} F^{\prime, \mu\nu} +
(D_\mu X)^\dagger D^\mu X ~, \label{eq:lkin}
\end{equation}
with the covariant derivative $D_\mu X \equiv (\partial_\mu - i \tilde{g} A^\prime_\mu) X$ and $\tilde{g}$ being the gauge coupling associated with the hidden gauge group.

The hidden scalar $X$ and the SM Higgs field $H$ have a common scalar potential. In unitary gauge we define $H^T=(0,h/\sqrt{2})$,
$X=x/\sqrt{2}$ with real scalar fields $h$ and $x$. The full scalar potential containing all terms of mass dimension 4 that are consistent with the symmetries of the model   reads
\begin{equation}
V= {1\over 4} \lambda_h h^4 + {1\over 4}\lambda_{hx} x^2 h^2 + {1\over 4}\lambda_x x^4 + 
{1\over 2} m_h^2 h^2 + {1\over 2} m_x^2 x^2 \;. \label{V}
\end{equation}
Here the real parameters
$\lambda_i$ and $m_i^2$  are the quartic couplings and mass terms, respectively. The scalar potential is such that not only $H$ but also $X$ develops a  vev:
\begin{equation}
\langle h \rangle =v,~~ \langle x \rangle =u ~.
\end{equation} 
Upon spontaneous  breaking of \uonep{} via $X$, the corresponding Goldstone boson is absorbed by the \uonep{} gauge field, leaving us with one real scalar degree of freedom and a massive vector field, which we will call $Z^\prime_\mu$ from now on. 

Using the minimisation conditions of the scalar potential and expanding both fields around their vevs, one finds the mass matrix of the scalar sector whose eigenvalues are given by 
\begin{equation}
m_{1,2}^2= \lambda_h v^2 + \lambda_x u^2 \mp 
\sqrt{(\lambda_x u^2 - \lambda_h v^2)^2 + \lambda_{hx}^2 u^2 v^2 }.
\label{eigenvalues}
\end{equation}
Defining the mass eigenstates by the following rotation
\begin{eqnarray}
 h_1 &=&   h \cos\theta  +  x \sin\theta \;, \nonumber\\
 h_2 &=&  h \sin\theta  -  x \cos\theta  \;, 
\end{eqnarray} 
the mixing angle $\theta$  is given by
\begin{equation}
\tan 2 \theta = {\lambda_{hx} u v  \over \lambda_h v^2 - \lambda_x u^2} \;.
\label{tan}
\end{equation}
The definition of $\theta$ is such that in the limit $\theta = 0$ the lighter of the two eigenstates is the SM--like Higgs boson\footnote{Note that this convention for $\theta$ differs from the one used in \cite{Lebedev:2011aq}.}. 
  
Setting the \uonep{} charge of $X$ to $+1$ we find
\begin{equation}
m_{Z^\prime}= \tilde g u \;, \label{eq:zprimemass}
\end{equation} 
and the following 
trilinear $SVV$ interactions relevant for our study
\begin{equation}
\Delta {\cal L} = {\tilde g}  m_{Z^\prime} \sin\theta
 ~h_1  Z^\prime_\mu Z^{\prime \mu} -
  {\tilde g}  m_{Z^\prime} \cos\theta
 ~h_2  Z^\prime_\mu Z^{\prime \mu} \;. \label{interaction}
\end{equation} 
These vertices are responsible for production of pairs of  $Z^\prime$s
at the LHC. 
\paragraph{Signatures of a $h  Z^\prime Z^{\prime}$ coupling:}
An important constraint on the model comes from the branching ratio of the invisible Higgs decays \cite{Djouadi:2011aa} which only applies if $m_{Z^\prime} < m_{1}/2$, that is if the hidden sector gauge boson is lighter than about 63 GeV. Bounds from the ``standard'' $Z^\prime$ searches do not apply to the  Higgsophilic case as they  assume couplings of the $Z^\prime$ to SM fermions and/or SM gauge bosons \cite{Langacker:2008yv}.

As outlined in the introduction, if the $Z^\prime$ is stable or decays into hidden sector states,
 the corresponding signature at a hadron collider would be missing transverse momentum which can, for instance, be  observed in conjunction with a jet.

\paragraph{Dark matter candidates:} 
In our framework, there exist several candidates for dark matter. A straightforward possibility is that the massive gauge fields themselves constitute DM.
In the Abelian case, only pairs of $Z^\prime$ couple to $h_1, h_2$, i.e. there exists a $\mathbb{Z}_2$--parity \cite{Lebedev:2011iq} which can be traced back to charge conjugation symmetry:
\begin{equation}
Z'_\mu \rightarrow - Z'_\mu ~.
\label{stabZ2}
\end{equation} 
This renders the $Z^\prime$ stable and  weakly coupled to the Standard Model, which are the prerequisites of viable DM candidates. The heavy $h_2$ limit of this set--up was studied in \cite{Lebedev:2011iq}, where it was found that all of the DM constraints can be satisfied for sub--TeV  $Z^\prime$ masses (see also \cite{Farzan:2012hh,Baek:2012se}). 
A recent analysis of the \uonep{} case can be found in \cite{Gross:2015cwa,Duch:2015jta}.

Another approach to the DM problem is to consider additional fields in the hidden sector that 
can account for DM. For instance, ``hidden fermions'' $\chi$ charged under \uonep{} can couple to the Higgsophilic gauge fields as follows
\begin{equation}
\Delta {\cal L} = \tilde g  ~ \overline{ \chi}  \gamma^\mu A^\prime_\mu \chi ~.
\end{equation}
In that case, after spontaneous symmetry breaking the  massive $Z^\prime$ can decay into fermionic DM $\chi$. In terms of collider phenomenology, this leads to the same missing $E_T$ signatures and hence would not change the results of this study. There are however differences in direct DM detection  as the hidden fermions have  loop--suppressed interactions with nucleons compared to those of the $Z^\prime$.

DM constraints on  our model depend on additional assumptions such as the  nature of  dark matter and its production mechanism(s) in the Early Universe. In this work, we  set these issues aside and focus exclusively on the collider aspects of our framework.

\subsection{Higgsophilic gauge bosons from \sunp{}}
\paragraph{ Abelian case parallel:}
The above considerations can straightforwardly be generalized to the non--Abelian case: Suppose we have an \sunp{} symmetry in the hidden sector instead of the \uonep{}. We now take $X$ to be an N-plet transforming in the fundamental representation of \sunp. The covariant derivative in Eq.~(\ref{eq:lkin}) then changes to
$ D_\mu = \partial_\mu -i \tilde g A^a_\mu T^a $, where $ A^a_\mu$  are the $N^2-1$
vector fields and $T^a $ are the  group generators satisfying\footnote{We use this normalization for easier translation from the Abelian case with charge +1. This differs from the normalization used in \cite{Gross:2015cwa} which is obtained by the replacement $\tilde g^2 \rightarrow \tilde g^2/4.$} Tr$(T^a T^b)=2 \delta^{ab}$. 

A vev of $X$ breaks   \sunp{} $\rightarrow$ \sunpm{}. In unitary gauge, $X$ is expressed as 
\begin{equation}
X= {1\over \sqrt{2}} 
\left( \begin{matrix}
0 \\
\vdots\\
0 \\
x 
\end{matrix} \right) 
\label{Nplet}
\end{equation}
with $x$ being a real scalar field  which gets a vev, analogously to the \uonep{} case. From the pattern of symmetry breaking, it is clear that 
$2N-1$ degrees of freedom of $X$ get absorbed and lead to $2N-1$ massive gauge
fields, while the remaining degree of freedom corresponds to the ``hidden sector
Higgs'' boson.  

In this gauge, the scalar potential is identical to that for the Abelian case (\ref{V}) and thus the conclusions that follow from that equation also apply. 
The only difference is that now $h_1$ and $h_2$ couple to $2N-1$ mass degenerate
bosons such that Eq.~(\ref{interaction}) now reads 
\begin{align}
\Delta {\cal L} = \sum_{i=1}^{2N-1}\Big( {\tilde g}  m_{Z^\prime} \sin\theta
 ~h_1  Z^\prime_{i,\mu} Z^{\prime \mu}_i -
  {\tilde g}  m_{Z^\prime} \cos\theta
 ~h_2  Z^\prime_{i,\mu} Z^{\prime \mu}_i \Big) \;. \label{interaction2}
\end{align}
Since $Z^\prime_{i, \mu}$ are indistinguishable experimentally, this effectively amounts
 to replacing 
\begin{equation}
\tilde g^2 \rightarrow   (2N-1) ~ \tilde g^2  
\label{rep}
\end{equation}
in cross sections and decay width calculations of the Abelian case. This expression bears resemblance to the 't Hooft coupling \cite{'tHooft:1973jz}  $\lambda= \tilde g^2 N$     for large $N$.

\paragraph{Residual \sunpm : }
At this stage, the other $(N-1)^2-1$ gauge bosons remain massless. As they have no coupling
to $h_{1,2}$  they do not play any role in our   collider analysis. However, they would affect the cosmological history of our Universe and thus necessitate further discussion.
 
In order to break the gauge group completely, one may invoke not just 1 but $N-1$ 
hidden sector Higgs fields $X_k$ in the fundamental representation of \sunp. 
When all of them get vevs, the symmetry gets  broken completely and all vector fields acquire mass.
 In general, all the remaining scalar degrees of freedom  mix independently with 
the Higgs, leading to a highly entangled scalar sector. However, one would not expect
all the mixings to be equally important. Hence, it is reasonable to make the  
 simplifying assumption that the mixing is dominated by one $N$--plet which we choose to be   the one in Eq.~(\ref{Nplet}). In this case, one may still neglect the
production of the remaining  $(N-1)^2-1$ gauge bosons at the LHC and the above result holds. 

Alternatively one could assume  condensation of \sunpm{} at low energies
which would make the relevant degrees of freedom massive (similarly to ``glueballs'' in QCD).

Either way we may focus on the couplings of
$h_1$ and $h_2$ to $2N-1$ massive vector bosons and ignore the rest. The only difference 
from the Abelian case would be the replacement in Eq.~(\ref{rep}). 

\paragraph{Dark matter candidates:} 
Consider for example $N=2$: As shown in \cite{Hambye:2008bq}, our considerations of
the Abelian gauge field DM equally apply to SU(2) as long as the symmetry 
is broken by a single SU(2) doublet. In this case,  the gauge fields couple 
to the physical scalars in pairs which renders the $Z^\prime_i$ stable.

Although the triple gauge vertex breaks an analog of the $\mathbb{Z}_2$--parity in Eq.~(\ref{stabZ2}), the interactions preserve a related $\mathbb{Z}_2 \times \mathbb{Z}_2$ symmetry,
\begin{eqnarray}
&& A^1_\mu \rightarrow - A^1_\mu ~~,~~ A^2_\mu \rightarrow - A^2_\mu  \;, \nonumber \\
&& A^1_\mu \rightarrow - A^1_\mu ~~,~~ A^3_\mu \rightarrow - A^3_\mu \;,
\end{eqnarray}
where the upper index refers to the SU(2) adjoint generators. This symmetry is sufficient to ensure stability of DM, while it actually generalizes to a custodial 
SO(3)  \cite{Hambye:2008bq}. Phenomenology of the SU(2) DM was studied in \cite{Hambye:2008bq}, see also \cite{Khoze:2014xha}.

The general \sunp{} case was analyzed in \cite{Gross:2015cwa}. It was found that the symmetries that stabilize DM include both inner and outer automorphisms of \sunp. These remain valid symmetries of the theory if CP is unbroken in the hidden sector. The resulting stable gauge fields are again viable DM candidates \cite{Gross:2015cwa}. 

As in  the Abelian case, there is the option  of having additional hidden sector fields $\chi$ charged under the \sunpm{} which could constitute dark matter.

\subsection{Perturbativity bounds}
We conclude this section  with a few words on the relevant theory limits.
   Requiring the hidden sector to be perturbative at the LHC energies 
   implies a bound on the 't Hooft coupling, 
 \begin{equation}
 \tilde g^2 N < 4 \pi^2 \;, \label{eq:pertbound}
 \end{equation}
where $4 \pi^2 $ represents the loop factor appearing at each  order in perturbation theory. The Abelian case emerges trivially from Eq.~(\ref{eq:pertbound}) by setting $N=1$. 

Further perturbativity constraints should be imposed on the scalar quartic couplings in Eq.~(\ref{V}). These however are relevant only if there is a significant hierarchy between the gauge boson mass and $m_2$ \cite{Gross:2015cwa} which we are not going to consider here.

\section{LHC monojet constraints}
\label{sec:monojet}

Constraints on the model depend strongly on the Higgsophilic gauge boson mass. If it is lighter than about 63 GeV, the SM--like scalar $h_1$ can decay into pairs of $Z^\prime$s. In that case, experimental constraints are rather strong and can be extracted from the results of \cite{Djouadi:2011aa}. For example, for $m_{Z^\prime} \sim 50$ GeV,
$\tilde g \sin\theta$ can be at most of order $10^{-2}$. A more recent analysis of this decay mode can be found in \cite{Chen:2015dea}.

In what follows, we therefore focus on the regime
\begin{equation}
m_1<2 m_{Z^\prime}<m_2 \;,
\end{equation} 
in which case the decay $h_2 \rightarrow Z^\prime Z^\prime$
is allowed and its width is enhanced by powers of $m_2/m_{Z^\prime}$
characteristic of the pseudo--Goldstone boson production. For heavier $Z^\prime$, the production cross section
is too small to have interesting constraints (see also \cite{Chen:2015dea}). Among other things, $Z^\prime$ production via off-shell Higgses
suffers from  destructive interference between the $h_1$ and $h_2$ contributions (see also \cite{Kauer:2015hia}). 

\begin{figure}[h!]
    \begin{center}
         \includegraphics[width=0.3\textwidth]{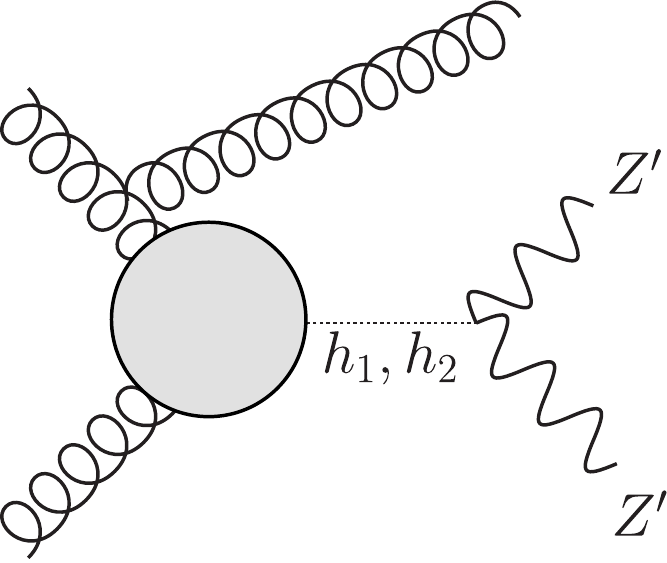}  
\vspace*{-2mm}
          \caption{\footnotesize Main contribution to the monojet production with missing $E_T$.
          }
\label{monojet}
\end{center}

\vspace*{-3mm}
\end{figure}

Here, we focus on the monojet signature shown in Fig.~\ref{monojet}. Other channels, such as Higgs production through vector boson fusion and  
$h_2$ visible decays, can  provide further important information about the model but are not discussed in this work. Related studies of fermion DM production have recently appeared in \cite{Khoze:2015sra}.

We consider $Z^\prime$ pair production in association with one hard jet via on--shell heavy Higgs production,
\begin{equation}
pp\rightarrow h_2\,j \rightarrow Z^\prime Z^\prime j,
\end{equation}
where $j$ denotes a parton level jet. The $Z^\prime$ will not be detected at the LHC and thus the corresponding signal is one large transverse momentum jet and 
large transverse missing momentum  which is back to back to the jet. Note that additional jets can arise from strong initial state radiation  and thus the above signal can be accompanied by further softer jets.

In the following, we first consider the invisible decay branching ratio for $h_2$. Then 
we  discuss constraints from the current monojet searches with 8 TeV LHC data and  the prospects at 14 TeV assuming an integrated luminosity of 600 fb$^{-1}$.

\subsection{BR($h_2\rightarrow$ invisible)}
When allowed kinematically, $h_2$  decays into SM particles, pairs of
$Z^\prime$ as well as pairs of $h_1$. Details of the relevant couplings and decay rate formulae for  $h_2 \rightarrow h_1 h_1$ can be found in \cite{Falkowski:2015iwa}.

The coupling between the light and heavy Higgses is given by
\begin{equation}
\Delta V = \frac{2 m_1^2+m_2^2}{2 v} \sin \theta \Big( \cos^2 \theta + \frac{v}{u} \sin \theta \cos \theta \Big) h_1^2 h_2.
\end{equation}
Here $v$ is the SM Higgs vev and $u$ is determined by the Higgsophilic gauge boson mass according to Eq.~(\ref{eq:zprimemass}).
The corresponding $h_2 \rightarrow h_1 h_1$ decay rate is then
\begin{equation}
\Gamma(h_2 \rightarrow h_1 h_1)= 
 \frac{(2 m_1^2+m_2^2)^2}{32 \pi  v^2 m_2^2} \sin^2 \theta \Big( \cos^2 \theta + \frac{v}{u} \sin \theta \cos \theta \Big)^2 \sqrt{1-\frac{4 m_1^2}{m_2^2}} ~. \label{eq:h2h1h1}
\end{equation}
From  Eq.~(\ref{interaction}) we find the decay width for   $h_2 \rightarrow Z' Z'$  to be
\begin{equation}
\Gamma(h_2 \rightarrow Z' Z')= \frac{\tilde g^2  \cos^2\theta  m_2^3}{32 \pi m_{Z'}^2 } \sqrt{1-\frac{4 m_{Z^\prime}^2}{m_2^2}} \left(1 - \frac{4 m_{Z^\prime}^2}{m_2^2} + \frac{12 m_{Z^\prime}^4}{m_2^4} \right)~. \label{eq:h2zpzp}
\end{equation}
Finally, the width of the $h_2$ decay into SM particles is obtained by rescaling the  heavy SM Higgs result,
\begin{equation}
\Gamma(h_2 \rightarrow \text{SM}) = \sin^2 \theta\ \Gamma_\text{SM}(m_h = m_2)~. \label{eq:h2sm}
\end{equation}
Eqs.~(\ref{eq:h2h1h1}-\ref{eq:h2sm}) determine the invisible decay branching ratio for $h_2$:
\begin{equation}
\text{BR($h_2\rightarrow$ invisible)} = \left(1 + \frac{\Gamma(h_2 \rightarrow h_1 h_1) + \Gamma(h_2 \rightarrow \text{SM})}{\Gamma(h_2 \rightarrow Z' Z')} \right)^{-1} ~. \label{BR}
\end{equation}
For small $\sin \theta$, the SM channels 
as well as $h_2 \rightarrow h_1 h_1$  are suppressed by $\sin^2 \theta$ and the $h_2 \rightarrow Z' Z'$ mode typically dominates. In addition,
the decay of $h_2$ into vector particles is enhanced by the usual  $ m_2/m_{Z'} $   
factor associated with the would--be Goldstone boson production.

To give an example,
for $m_2 = 300$ GeV, the SM heavy Higgs width is $\Gamma_{\rm SM} \simeq 10 $ GeV.
Then for $\tilde g = 1$, $m_{Z'} = 100$ GeV and $\sin\theta=0.4$, the $h_2$ invisible decay branching ratio exceeds 90\%.

\subsection{Current constraints from the LHC at 8 TeV}
The ATLAS and CMS collaborations have presented limits on invisible Higgs decays using 20.3 fb$^{-1}$ of data at $\sqrt{s}=8$ TeV \cite{Aad:2015zva,Chatrchyan:2014tja,ATLAS:2015yda,ATLAS:2015yda,Aad:2014iia,Aad:2015uga}. No excess above the SM has been observed and CMS and ATLAS have derived 95\% C.L. limits on the production cross section times the branching ratio as a function of the Higgs-like boson mass. We assume that the production mechanism in our scenario is the same as that in the SM. However, the additional factor of the mixing between the SM-like Higgs and the singlet heavily suppresses the production rate and thus no limits on our model can be derived from 8 TeV data. 

\subsection{Future limits from the LHC at 14 TeV}
In this subsection, we discuss prospects of constraining our Higgs portal model at the LHC with $\sqrt{s}=14$ TeV. We extrapolate an existing ATLAS monojet search at 8 TeV to 14 TeV adding signal regions with stricter cuts on the transverse missing energy and the leading jet but without optimizing the selection cuts.
\paragraph{Simulation:}
While the more recent monojet study \cite{Aad:2015zva} puts limits on an invisibly decaying SM--like Higgs boson,   for our 14 TeV projection  we have closely followed the slightly older monojet search of Ref.~\cite{Aad:2014nra}.
By the time Ref.~\cite{Aad:2015zva} was published the SM backgrounds for this study had already been fully simulated based on Ref. \cite{Aad:2014nra}. The simulation of SM backgrounds with large cross sections requires large computer resources and since both studies have  similar selection cuts (apart from the details of the jet veto) we have decided to adhere to Ref.~\cite{Aad:2014nra} in this work.

We have implemented the relevant kinematic selection cuts for the signal regions in this study. All monojet signal regions demand a lepton veto and a maximum of three jets with $p_T>30$ GeV. An additional requirement is imposed on the azimuthal angle between the missing transverse vector and the jets, $\Delta\phi(\rm{jet},p_T^{\rm miss})>0.4$, in order to suppress the QCD multijet background. Finally, five signal regions M1, M2, M3, M4 and M5 are defined with increasing cuts on the transverse momentum of the leading jet and the total missing transverse energy of the event. They are listed in Table \ref{tbl:selection}.

\begin{table}
\centering
\begin{tabular}{l||c|c|c|c|c}
\hline
\hline
Cut & M1 & M2 & M3 & M4 & M5 \\
\hline
lepton veto & \multicolumn{5}{c}{\text{yes}} \\
$N_j(p_T > $ 30 GeV$, |\eta| < 2.8)$ &  \multicolumn{5}{c}{$\leq 3$}\\
$\Delta\phi(\vec{p}_{\rm{jet}},\vec{p}_T^{\rm miss})$ & \multicolumn{5}{c}{$>0.4$} \\
\hline
$p_T(\text{leading jet})$ in GeV & $\geq 280$ & $\geq 320$ & $\geq 450$ & $\geq 500$ & $\geq 550$ \\
$E_T^{\text{miss}}$ in GeV & $\geq 220$ & $\geq 320$ & $\geq 450$ & $\geq 500$ & $\geq 550$ \\
\hline
\hline
\end{tabular}
\caption{Selection cuts used for the $\sqrt{s} = 14 $ TeV monojet analysis.}
\label{tbl:selection}
\end{table}

We have generated the parton level signal events within the {\tt POWHEG2} framework \cite{Nason:2004rx,Frixione:2007vw,Alioli:2010xd} which then have been passed to {\tt Pythia6.4} \cite{Sjostrand:2006za}. We have produced $gg\rightarrow h_2$ \cite{Bagnaschi:2011tu} and $VV\rightarrow h_2$ \cite{Nason:2009ai} samples. Since the $Vh_2$ production mechanism is subdominant, we have omitted the production channel of $h_2$ in association with a gauge boson. The $gg\rightarrow h_2$ sample dominates the total production cross section. The cross sections for the various signal production modes have been taken from \cite{ATLAS_higgs_cross}. The signal event generation has been validated against the results on invisible Higgs decays from \cite{Aad:2015zva} before generating signal events for the 14 TeV study.

Let us  briefly discuss the major SM backgrounds for the 14 TeV study.
The main background is  the $Z$ boson production in association with one jet where the $Z$ decays into a pair of neutrinos. The $W j$ production with the $W$ decaying leptonically also contributes significantly, most importantly via the decay into a tau and a neutrino. $t \bar t$ events give a small contribution but are important for choosing the cuts for the signal regions. We omit the single top background since its cross section is by a factor of 4 smaller than the $t\bar t$ background which itself only contributes at the percent level. For the same reason, we have neglected the $Z/\gamma^* j$ and SM diboson as well as dijet/trijet QCD backgrounds.

We estimate the dominant SM backgrounds as follows: The  $W j$ and $Z j$ backgrounds are generated with {\tt Sherpa2.1.1} \cite{Gleisberg:2008ta} including up to 3 partons with {\tt CTEQ10} PDF \cite{Lai:2010vv}. The $t \bar t$ background has been simulated with {\tt POWHEG2} \cite{Frixione:2007nw} and the parton level events were passed to {\tt Pythia6.4.25} \cite{Sjostrand:2006za} with {\tt CTEQ6L1} parton distribution function \cite{Pumplin:2002vw}. The $t\bar t$ cross section has been determined with {\tt Top++2.0} \cite{Czakon:2011xx}. 

Our 14 TeV monojet analysis has been implemented  into the {\tt CheckMATE1.2.1} framework \cite{Drees:2013wra}. {\tt CheckMATE} uses the fast detector simulation {\tt Delphes3.10} \cite{deFavereau:2013fsa} with heavily modified detector tunings of the ATLAS detector. For a given event sample, it determines the number of expected signal events passing the selection requirements. Its \emph{AnalysisManager} feature allows for an easy implementation of new studies \cite{Kim:2015wza}. We have used \emph{AnalysisManager}   to implement the aforementioned selection cuts and obtain the expected background numbers. The latter is used by {\tt CheckMATE} to automatically calculate the CL$_\text{S}$ value \cite{Read:2002hq} in order  to quantify the compatibility of the signal prediction with the observation, which for our prospective study equals the SM expectation. The statistical errors of the signal and of the background are taken into account. In addition, we assume a 10$\%$ theory  error on the signal. The overall magnitude of the systematic errors and its relative contribution to the statistical uncertainty is hard to estimate for a future high luminosity LHC run. We therefore determine optimistic limits for negligible systematic errors and discuss their impact on the final result at the end of the section.

\begin{table}
\centering
\begin{tabular}{l||r|r|r|r|r|r}
\hline
\hline
SR &  $Z j$& $W j$ & $t \bar t$ & total & signal&$S/\sqrt{B}$\\
\hline
M1 & 2378934 & 2024466 & 67821 & 4471221 & 13268& 6.3\\
M2 & 742710 & 442296 & 13327 & 1198333 & 4894& 4.5\\
M3 & 207804 & 102852 & 2656 &  313312 & 1514& 2.7\\
M4 & 80730 & 30036 & 1118 & 111884 & 942& 2.8 \\
M5 & 33252 & 11610 & 625 & 45487 & 594& 2.8\\
\hline
\hline
\end{tabular}
\caption{Number of background and example signal events ($m_2=200$ GeV, $\sin\theta=0.4$ and \mbox{BR($h_2 \rightarrow$ invisible) $= 1$)} in the signal regions M1 to M5 at the LHC with $\sqrt{s} = 14 $ TeV and an integrated luminosity of 600 fb$^{-1}$. In the last column, we have estimated the statistical significance with $S/\sqrt{B}$.}
\label{tbl:background_numbers}
\end{table}

\paragraph{Benchmark Parameters:}
To allow for a large phase space and a large coupling, we set \ $m_{Z^\prime}=65$ GeV in our benchmak study.  Eq.~(\ref{BR}) implies that the result depends on the combination
$\tilde g/m_{Z^\prime}$ as long as $m_2^2 \gg m_{Z^\prime}^2$, therefore our bounds for other values of $m_{Z^\prime}$ can be obtained
by an appropriate rescaling of $\tilde g$. For heavier $Z^\prime$, the kinematic suppression factor must also be taken into account.

Obviously, the production cross section is sensitive to $\sin\theta$, for which we take two representative values, $\sin\theta=0.3,0.4$. The LHC constraints on $\sin\theta$ from invisible decays of a heavy Higgs depend on $m_2$ as well as 
BR$(h_2\rightarrow\,\rm{invisible})$ since these searches are based on visible final states, e.g. photons and leptons. For large BR$(h_2\rightarrow\,\rm{invisible})$, the usual bounds \cite{Falkowski:2015iwa}  relax and the above values for $\sin\theta$ are
consistent with the data.

\begin{figure}
\begin{center}
\includegraphics[width=0.45\textwidth]{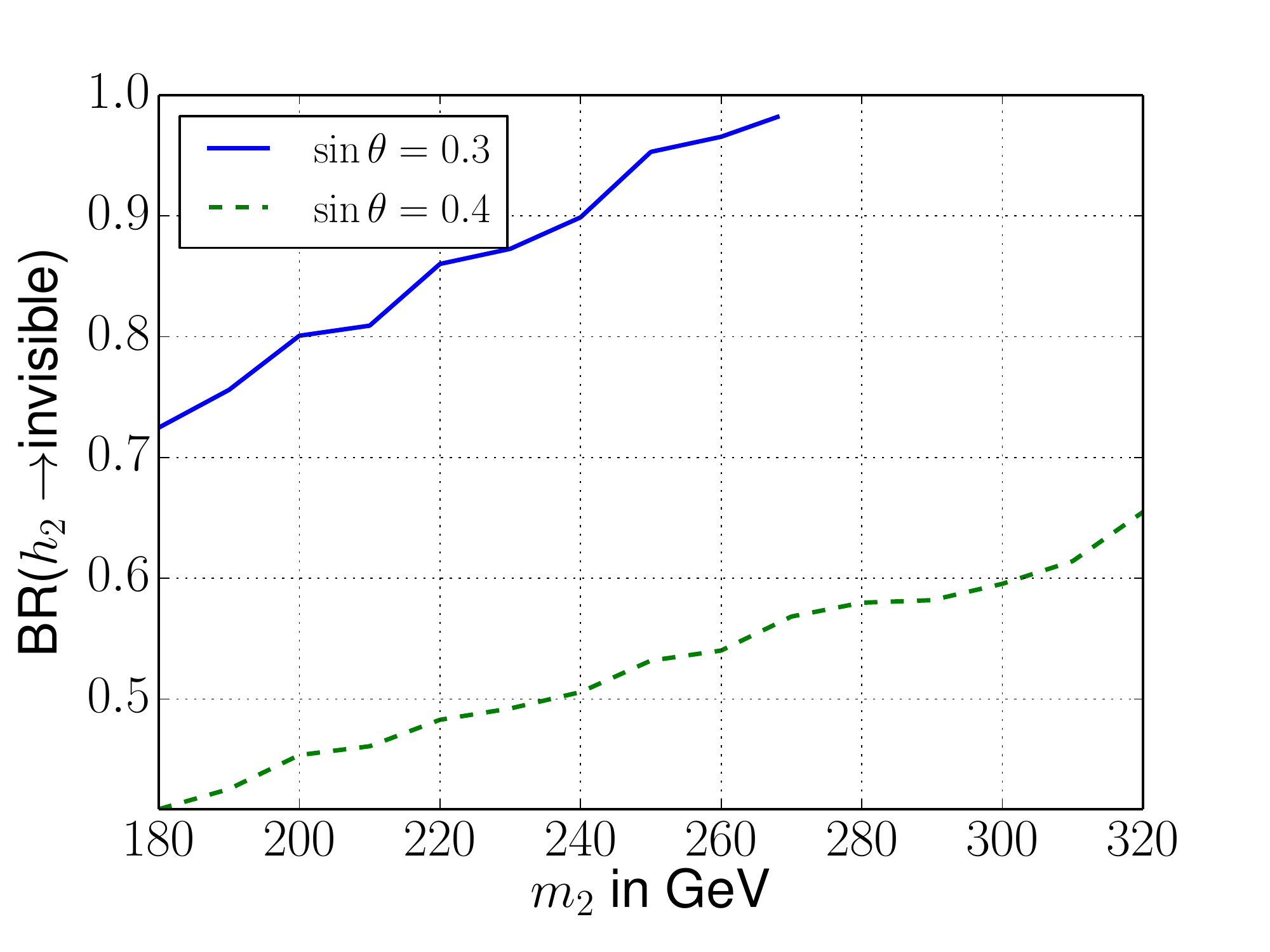} \qquad
\includegraphics[width=0.45\textwidth]{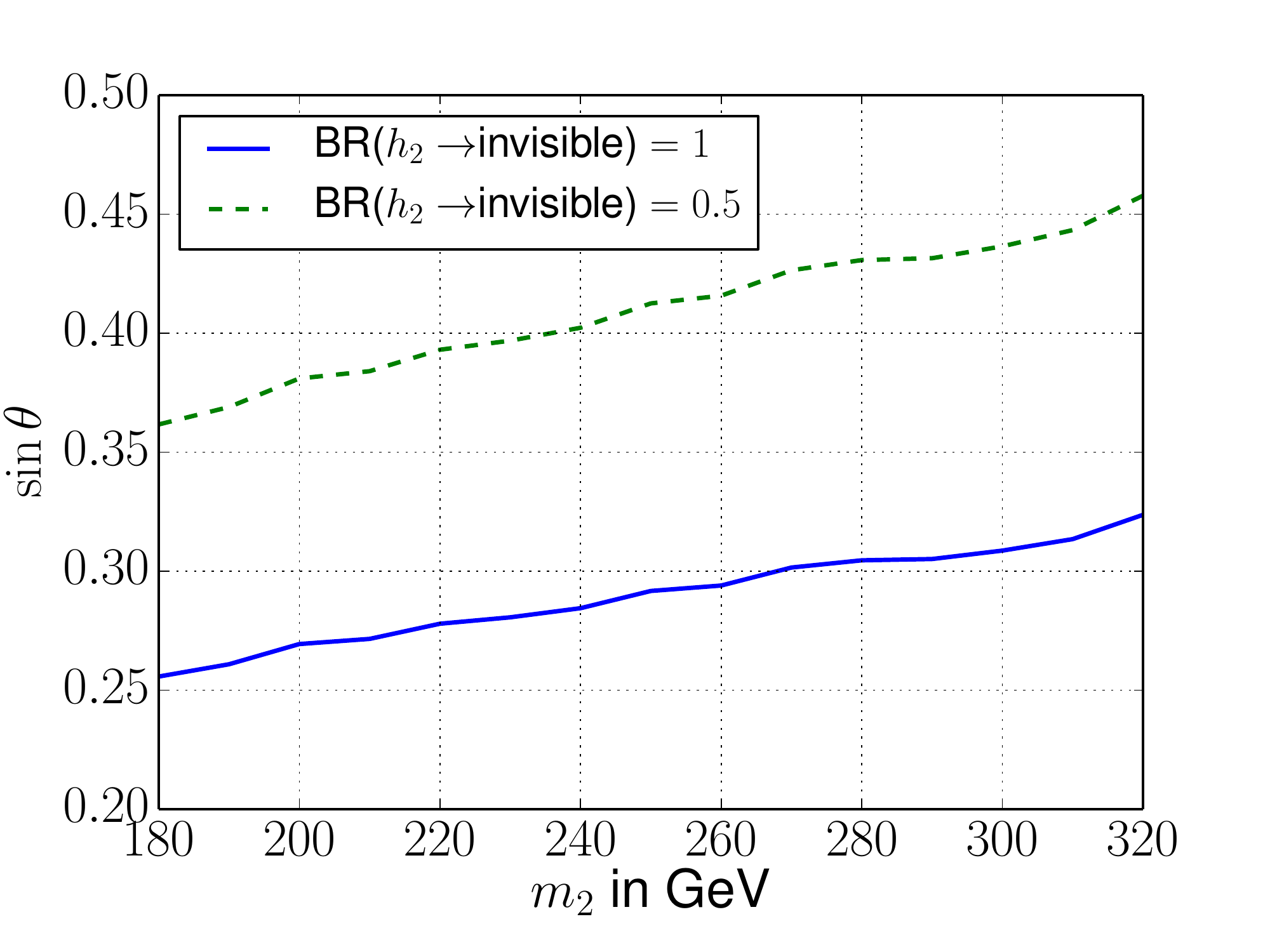} \\
\vspace{-0.75cm}
a) \hspace{0.45\textwidth} b)\hspace{0.45\textwidth}
\caption{Expected future 95\% C.L. limits on the heavy Higgs boson at $\sqrt s = 14$ TeV with an integrated luminosity of 600 fb$^{-1}$. a) Upper limits on BR($h_2 \rightarrow$ invisible) for fixed mixing angles. b) Upper limits on the mixing angle for fixed $h_2$ invisible decay branching ratios.}
\label{fig:exclusion_mass_BR}
\end{center}
\end{figure}

\paragraph{Results:}
In Table~\ref{tbl:background_numbers}, we list the total number of $Z+j$, $W+j$, $t\bar t$ and   the sum of the background events for a signal benchmark point  $m_2=200$ GeV, $\sin\theta=0.4$ and BR($h_2\rightarrow$invisible)=1 at the LHC with $\sqrt{s}=14$ TeV for an integrated luminosity of 600 fb$^{-1}$. In our numerical studies, 
we  determine the 95$\%$ CL$_\text{S}$ limits. However, for illustration,  in the last column we show the statistical significance of the signal estimated with $S/\sqrt{B}$, where
 $S$ and $B$ are the number of signal and background events, respectively. 
The $Z j$ and $W j$ production are the dominant SM backgrounds as  expected. 
The $t\bar t$ background is heavily reduced by the jet veto \cite{Drees:2012dd} which makes it negligible. We obtain the best statistical significance, namely above six, in the signal region M1. One should keep in mind however that no systematic error has been included.

In Fig.~\ref{fig:exclusion_mass_BR}a), we present our   95$\%$ C.L exclusion limits for BR$(h_2\rightarrow\,\rm{invisible})$ as a function of the $h_2$ mass for  an integrated luminosity of 600 fb$^{-1}$. In this mass range, the LHC would be sensitive to invisible decay branching ratios down to 40\% for $\sin \theta = 0.4$.
 Heavier $h_2$ weaken the limits and lead to the sensitivity range determined by the point at which BR$_\text{limit} = 1$. As can be seen in Fig.~\ref{fig:exclusion_mass_BR}a),   at  $\sin \theta=0.3$ the monojet signal can be useful for $m_2$ up to  270 GeV. 

The monojet bounds could also be interpreted from a different angle.
One may assume that the hidden gauge coupling is large such that
BR$(h_2\rightarrow\,\rm{invisible})\simeq 1$. In this case, one would
instead get a bound on $\sin\theta$ as shown in  
Fig.~\ref{fig:exclusion_mass_BR}b).
Clearly, there are also other probes of $\sin\theta$ such as the 
$h_1$ couplings to matter which will likely set a stronger bound.
When $\sin\theta$ is determined and the $h_2$ resonance is found, the interpretation of the monojet signal in terms of $\tilde g/m_{Z^\prime}$ becomes unambiguous.


  \begin{figure}
\begin{center}
\includegraphics[width=0.4\textwidth]{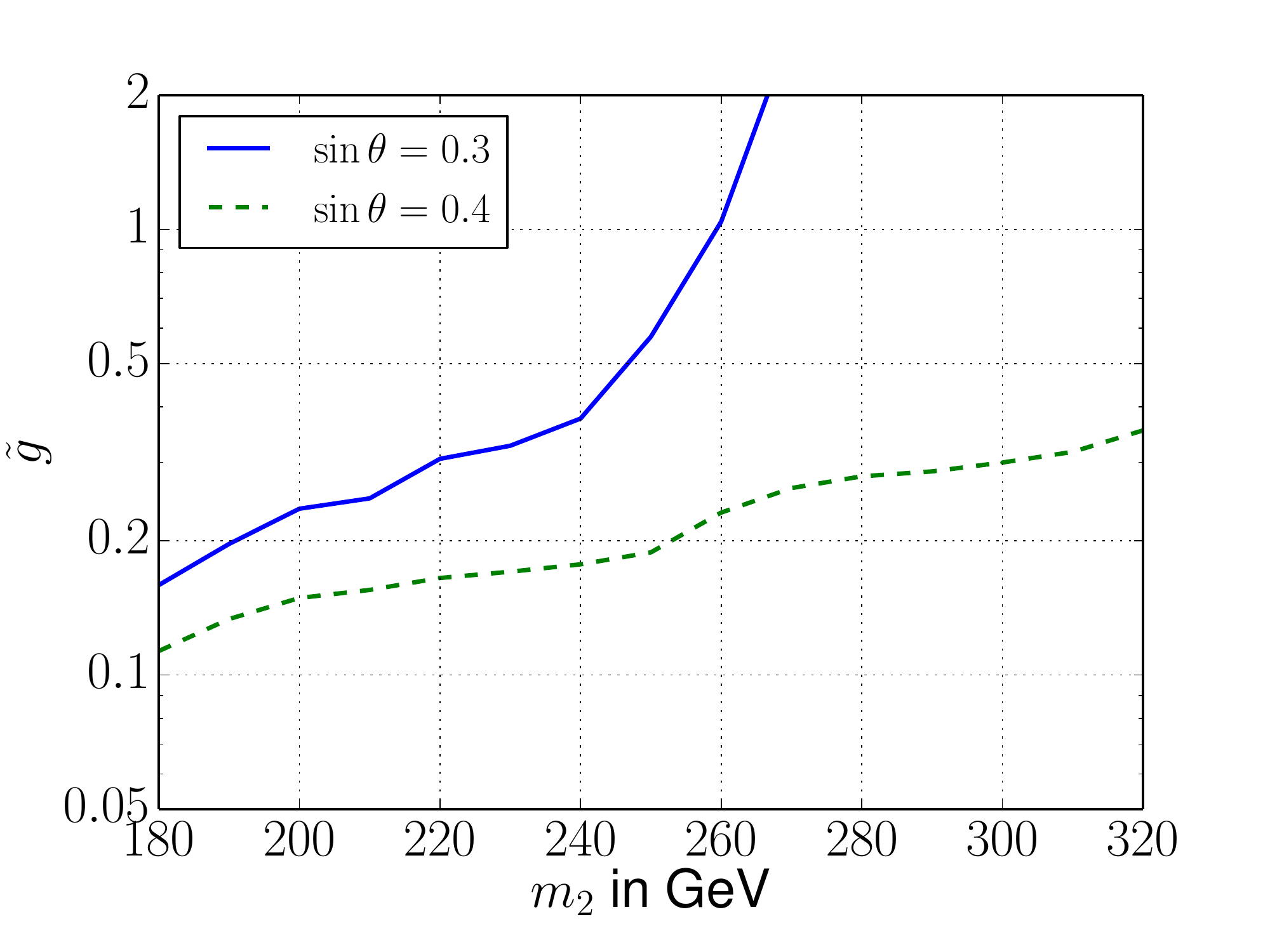} \qquad
\includegraphics[width=0.4\textwidth]{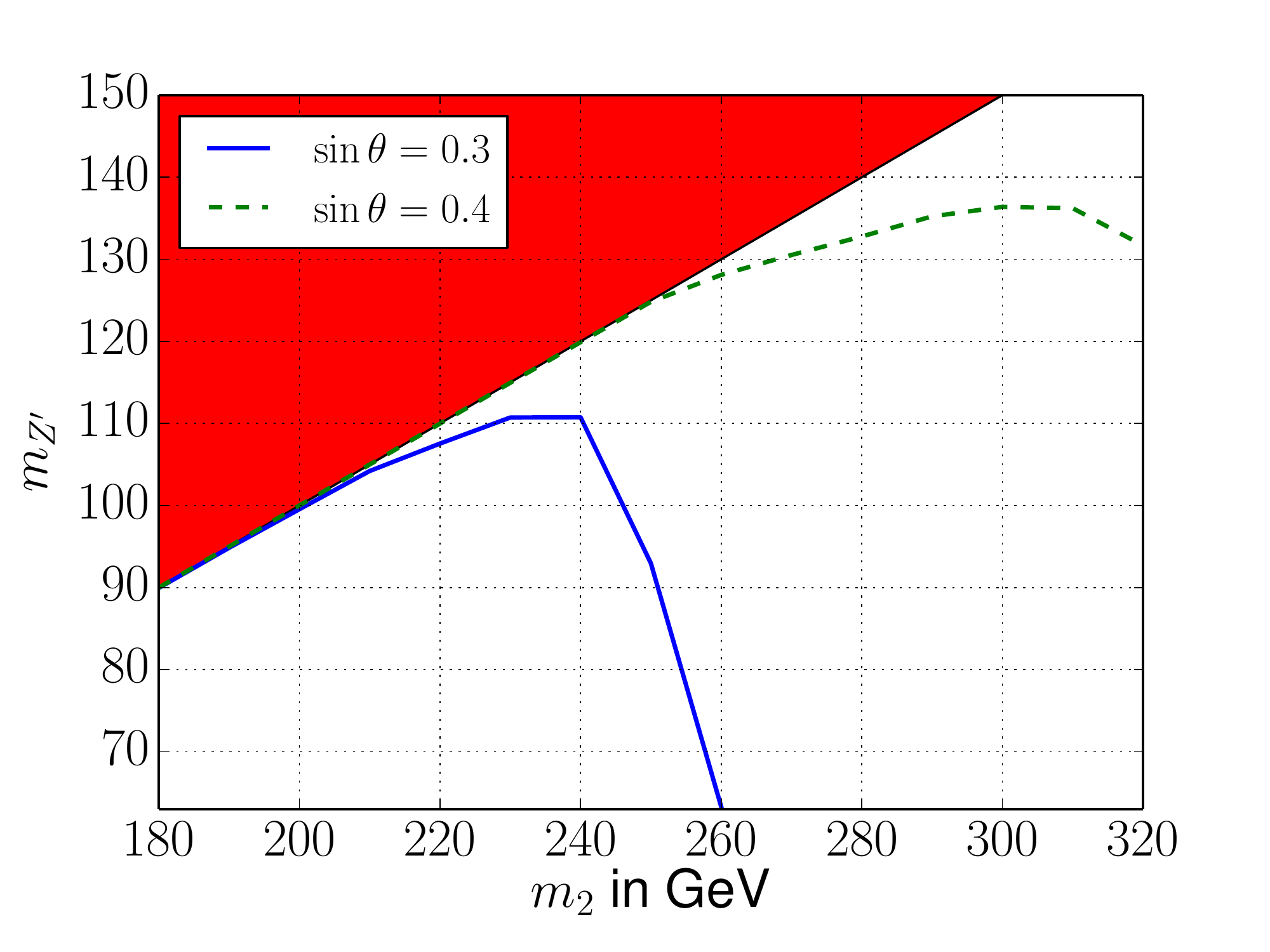} \\
\vspace{-0.75cm}
a) \hspace{0.45\textwidth} b)\hspace{0.45\textwidth}
\caption{95\% C.L. future limits on the hidden gauge sector parameters  at $\sqrt s = 14$ TeV with an integrated luminosity of 600 fb$^{-1}$ and two representative values of $\sin \theta$. a) Upper limit on the hidden sector gauge coupling $\tilde{g}$ for $M_{Z^\prime} = 65$ GeV. b) Lower Limit on the hidden sector gauge boson mass for $\tilde{g} = 1$. The red shaded area does not allow for a decay into two on-shell $Z^\prime$ bosons.}
\label{fig:coupling}
\end{center}
\end{figure}
  
The limits on the branching ratio are translated into the upper bounds on the coupling constant $\tilde g$ in Fig.~\ref{fig:coupling}a), for the two chosen values of $\sin\theta$. Interestingly, $\tilde g$ as small as $10^{-1}$ could in principle be constrained.  If, on the other hand,
one assumes a large $\tilde g$, BR($h_2\rightarrow$ invisible) probes a
wide range of $m_{Z^\prime}$, as shown in Fig.~\ref{fig:coupling}b), which can cover the entire kinematic reach
$m_{Z^\prime} \leq m_2/2$."   
 
 As explained in previous sections, the above limits can   be reinterpreted in the non--Abelian case by taking into account the multiplicity factor in Eq.~(\ref{rep}).

In our study, the signal to background ratio is roughly $S/B\lesssim1\%$. The shape of the signal and the dominant irreducible background process $Z(\rightarrow \nu\bar\nu)+j$ as well as $W(\rightarrow \tau\bar\nu)+j$ are very similar which makes the signal extraction extremely difficult. In the above considerations, we have not included the systematic uncertainty. This is a limiting factor in our study since one expects  a tangible systematic error on the background. The $Z(\rightarrow\nu\bar\nu)+j$ background can be  determined directly from data. One can measure the rate of $Z(\rightarrow \ell\ell)+j$ with the $Z$ decaying into electron or muon pairs. The $Z(\rightarrow\nu\bar\nu)+j$ cross section can be calculated from the known $Z$ branching ratios. However, the statistical fluctuation will still be  too large since $\sum_{\ell=e,\mu}\text{BR}(Z\rightarrow\ell\ell)\approx \sum_{i=e,\mu,\tau}\text{BR}(Z\rightarrow\nu_i\bar\nu_i)/3$. Actually it turns out that the $Z(\rightarrow \ell\ell)+j$ sample is by a factor of 5.3 smaller than that of $Z(\rightarrow\nu\bar\nu)+j$ in the signal region if detector effects are included. The resulting error of the $Z(\rightarrow\nu\bar\nu)+j$ background is thus $\sqrt{5.3}\approx2.3$ times larger than the statistical error. Unless the luminosity is very high, the data driven method will not reduce sufficiently  the total background uncertainties \cite{Drees:2012dd,Vacavant:2000wz}.

In addition, the $W j$ background  has a non--negligible systematic error. If one takes the systematic error into account, the signal regions with a harder kinematic cut on the leading jet as well as on the missing transverse momentum perform better than  the M1 signal region. In addition, the signal to background ratio  improves slightly. However, due to a significant loss of signal events, large integrated luminosities will be required and even then, only small parts of the parameter space will be covered. In practice, we find that the monojet signal is a useful probe of our model if the background is known to within less than one percent. Such a level of precision would be
challenging to achieve, yet one should not discount possible developments on the 
experimental side.

\section{Conclusion}

We have considered the possibility that the Higgs field serves as a portal into a hidden sector endowed with gauge symmetry. Due to the mixing with the hidden ``Higgs'', the 125 GeV scalar observed at the LHC is the only SM particle that couples to the hidden gauge bosons. The latter could  either  be stable or decay invisibly. If these are 
sufficiently light, the scenario is already constrained by the Higgs invisible decay.

In this work, we have explored a monojet signature of the Higgsophilic gauge bosons.
If these are heavier than about 63 GeV but below half the mass of the heavy  ``Higgs'' $h_2$,
they can be produced through on--shell decays of $h_2$. We find that the statistics allow
one to probe invisible decay  branching ratios of the heavier Higgs down to $40 \%$, or, in other terms, the hidden sector gauge coupling down to $10^{-1}$. Systematic uncertainties are a limiting factor which must be reduced to within one percent in order to gain the required sensitivity. This also implies that other channels with potentially lower systematic uncertainties, such as vector boson fusion, should be explored in more detail.

\section*{Acknowledgements}
\noindent 
This research was supported by the Munich Institute for Astro- and Particle Physics (MIAPP) of the DFG cluster of excellence ``Origin and Structure of the Universe''. 
The work of J.S. Kim has been partially supported by the MINECO, Spain, under contract FPA2013-44773-P; Consolider--Ingenio CPAN CSD2007-00042 and the Spanish MINECO Centro de excelencia Severo Ochoa Program under grant SEV-2012-0249. 
O.L. acknowledges support from the Academy of Finland project ``The Higgs boson and the Cosmos''.

\end{document}